\documentclass[aps,prb,reprint]{revtex4-2}
\usepackage{amssymb} 
\usepackage{graphicx}
\usepackage{dcolumn}
\usepackage{xcolor}
\usepackage{bm}
\usepackage{setspace}
\usepackage{hyperref}
\usepackage{orcidlink}
\usepackage{float}  

\usepackage{array}

\begin{document}

\preprint{APS/123-QED}

\title{Electron Temperature and Electron-Electron Scattering Length in GaAs/AlGaAs Using Mesoscopic Multiparallel Aperture Geometries}

\author{Arati Thapa\,\orcidlink{0000-0002-6138-8075}}
\email{aratithapa@vt.edu}
\author{Rishav Khatiwada\,\orcidlink{0009-0003-6038-6267}}
\author{Prakash Sharma\,\orcidlink{0009-0002-1117-0956}}
\author{Jean J. Heremans\,\orcidlink{0000-0002-6346-8597}}
\email{heremans@vt.edu}
\affiliation{Department of Physics, Virginia Tech, Blacksburg, Virginia 24061, USA}

\author{Adbhut Gupta}
\author{Kirk W. Baldwin}
\author{Loren N. Pfeiffer}
\author{Mansour Shayegan}
\affiliation{Department of Electrical Engineering, Princeton University, Princeton, New Jersey 08544, USA}

\begin{abstract}
Quantifying electron temperatures and electron–electron scattering lengths is essential for understanding electron transport regimes in two-dimensional systems. To that end, this work experimentally investigates low-temperature ballistic transport using semiclassical cyclotron orbits in a multiparallel aperture geometry. The mesoscopic geometry is fabricated on a GaAs/AlGaAs heterostructure of very high electron mobility. The amplitudes of maxima in magnetoresistance due to ballistic transport are observed to decay exponentially with temperature, allowing determination of the electron–electron scattering length from the data. A quantitative relation between electron temperature and an applied dc bias heating current is established by comparing magnetoresistance peak amplitudes measured as a function of temperature and dc bias, reflecting bias-induced Joule heating. In addition, the resistance measured near zero magnetic field decreases with increasing temperature, providing signatures of superballistic conductance.

\end{abstract}

\maketitle


\section{Introduction}

Electron-electron (e-e) scattering has an important role in solid-state electron systems, among other phenomena affecting the fundamental quantity of quasiparticle lifetime \cite{zheng1996, Giuliani, Polini}, inducing localization effects in electronic transport \cite{Taboryski, Tsui}, and controlling the hydrodynamic electron transport regime \cite{Gurzi_95, De_Jong, Gusev, Alekseev, Scaffidi,Scaffidi_24}. The e-e scattering process strongly depends on electron temperature $T_e$, with the e-e scattering length (or the quasiparticle lifetime) decreasing approximately as $1/T_e^2$ \cite{Adbhut2021,gramila1991mutual, narozhny2022hydrodynamic,Sarypov}. A change in $T_e$, which can be achieved by Joule heating via an applied current through the electron system, will vary e-e scattering and hence can form a parameter to shape electronic properties. The relevant current densities are readily reached in mesoscopic geometries and moreover e-e scattering has a particular impact on transport in mesoscopic systems \cite{Adbhut2021, Gupta_21}. Hence, mesoscopic electronic transport phenomena sensitive to e-e scattering can serve simultaneously to quantify e-e scattering \cite{Adbhut2021, Egorov, keser2021}, to quantify $T_e$, to change $T_e$ by current-induced heating, and to study the effects of e-e scattering on transport. We report here on the effect of electron heating in mesoscopic non-equilibrium electron transport in an ultra-high-mobility two dimensional electron system (2DES) in a GaAs/AlGaAs heterostructure, where $T_e$ is increased by Joule heating via a variable dc bias current, $I_{\mathrm{DC}}$~\cite{Dzurak_1994,Huang, Wang1,wang2,Molenkamp,Molenkamp_1994,Williamson,jain2026heating, Rojek, Cumming}. The quantification of the relation between $T_e$ and $I_{\mathrm{DC}}$ is necessary to identify the role of effective $T_e$ on mesoscopic transport. 

In our study, we extract $T_e$ from current-induced modifications of ballistic magnetoresistance maxima, enabling a determination of the $T_e$-$I_\mathrm{DC}$ relation that does not rely on thermoelectric effects \cite{Molenkamp,Molenkamp_1991, Molenkamp_Vol65, Molenkamp_1994,jain2026heating} or noise thermometry \cite{kurdak1995electron, huard2007electron}. We quantify the correlations between the e-e scattering length, $I_\mathrm{DC}$ and $T_e$. The method serves as a sensitive electron calorimeter that directly probes the thermal dynamics and properties of hot electrons in mesoscopic systems. In this method, electron heating decreases the e-e scattering length which directly and measurably modifies the ballistic magnetoresistance. 

The present work employs an approach akin to transverse magnetic focusing (TMF) ~\cite{Adbhut2021, Peide, Nakazato, Egorov, Hornsey_1993, Williamson, Heremans_TMF}, a powerful technique for studying ballistic transport, used here in an uncommon but compelling geometry. As explained below, we use a multiparallel aperture geometry ~\cite{Nihey_1990, F.Nihey_1990} which offers several advantages over the TMF geometry. Whereas the TMF geometry requires a 4-point nonlocal magnetoresistance measurement, the multiparallel aperture geometry is used in a simpler local magnetoresistance measurement over a perforated barrier (in principle 2-point, but implemented in 4-point to avoid contact resistances). As a consequence, the magnetoresistance is symmetric in the applied magnetic field, unlike for TMF. Several multiparallel aperture geometries can readily be connected in series, allowing experimental flexibility. Additionally, the geometries take little sample space. The multiparallel aperture geometry allows a direct measurement of the resistance of the channels perforating the barrier. This feature is used here to study hydrodynamic electron transport effects in the same narrow channels where $T_e$ and e-e scattering are characterized, and the feature allows a study of hydrodynamic and ballistic transport in the same structure. Further, the multiparallel aperture geometry naturally averages the magnetoresistance over several substructures arranged in parallel, mitigating artefacts arising from imperfections in individual mesoscopic structures. 

In ultrapure materials, such as high-mobility GaAs 2DESs~\cite{SarmaPRB106-2022} and graphene, momentum relaxation to the lattice by impurity scattering is strongly suppressed. At low temperatures ($\lesssim 15$ K) relaxation by electron-phonon scattering is suppressed as well. The e-e scattering then provides the dominant scattering mechanism~\cite{Gurzhi, Scaffidi, Gupta_21}. In 2DESs with very long momentum-relaxing (MR) scattering length (mobility mean-free path) $\ell_{\mathrm{MR}}$, the amplitude of ballistic transport signals is thus dominantly affected by e-e scattering. As e-e scattering conserves the total momentum of the 2DES, we denote the e-e scattering length as the momentum-conserving (MC) length $\ell_{\mathrm{MC}}$. The present experiments provide a means of quantifying $\ell_{\mathrm{MC}}$ and $T_e$. 
In measurements where the sample temperature was varied while $I_{\mathrm{DC}}=0$, we denote the lattice temperature as $T$ and we assume $T_e=T$ to a good approximation. In contrast, current-dependent measurements involve $I_{\mathrm{DC}}\neq 0$, where Joule heating causes $T_e>T$. In such measurements, we denote the electron temperature explicitly as $T_e$.
\section{ Methods and Materials}

\begin{figure*}[htbp]
    \centering
    \includegraphics[height=0.3\textheight]{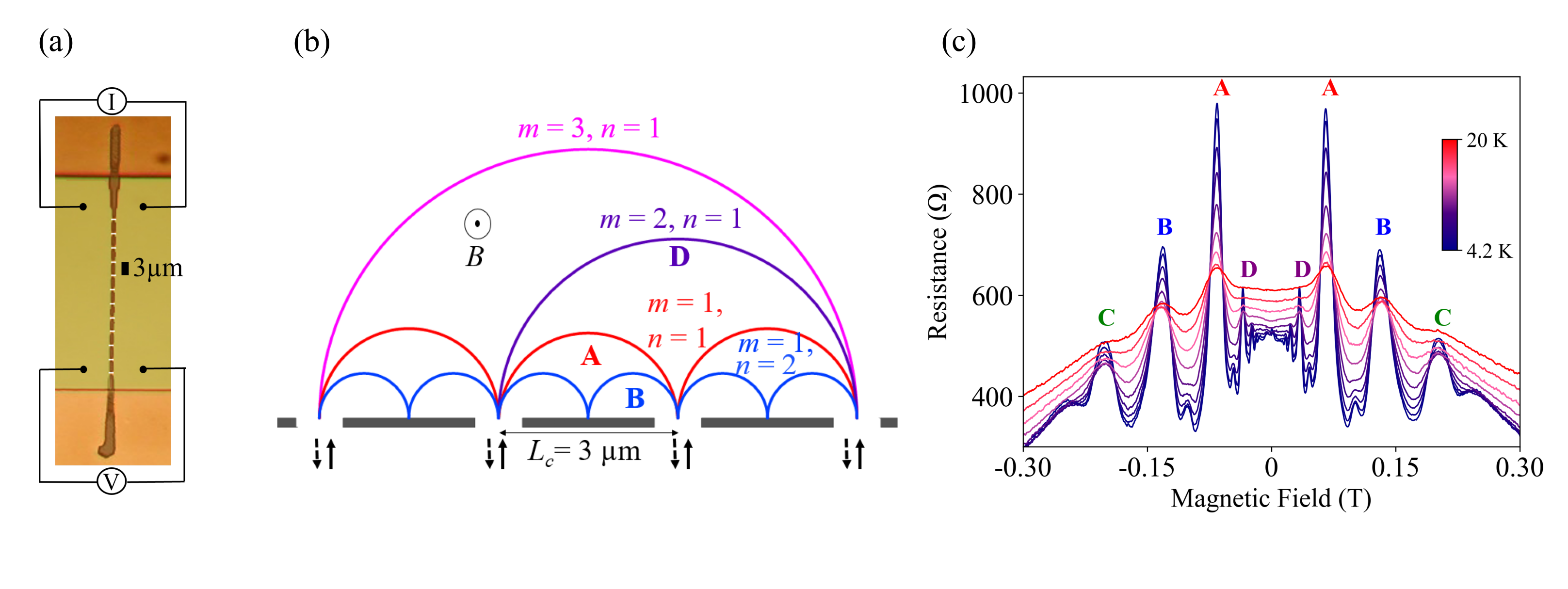}
    \caption{(a)  Optical micrograph of the 12-aperture perforated-barrier geometry showing the dimensions and the 4-point resistance measurement configuration with current and voltage contacts marked. (b) Schematic representation of selected cyclotron orbits corresponding to different $m$ and $n$. (c) Magnetoresistance measurements parametrized in $T$, showing maxima corresponding to different combinations of $m$ and $n$ indicated in (b). Peak \textbf{A} corresponds to $m=1, \, n=1$; peak \textbf{B} to $m=1, \, n=2$; peak \textbf{C} to $m=1, \, n=3$; and peak \textbf{D} to $m=2, \, n=1$. The resistance values correspond to a single aperture.}
    \label{fig1}
\end{figure*}

We devised a multiparallel aperture geometry (Fig.~\ref{fig1}(a)) ~\cite{Nihey_1990, F.Nihey_1990} on a GaAs/AlGaAs ultra high mobility 2DES to extract $\ell_{\mathrm{MC}}$ and change $T_e$. The perforated-barrier geometry enables us to resolve ballistic transport features, from which $\ell_{\mathrm{MC}}$ can be quantitatively determined. The GaAs/AlGaAs heterostructure was grown by molecular beam epitaxy \cite{Chung, Chung_2022}, and features a 30 nm wide quantum well where the 2DES resides, at a depth of 212 nm. After LED illumination, the 2DES has at $T$ = 4.2~K a 2D electron density $N_s = 3.66\times10^{15}\,\mathrm {m}^{-2}$ and mobility $ \mu =  920\,\mathrm{m}^2/\mathrm{V\,s}$. At $T$ = 4.2~K we then obtain 2D resistivity $R_{\square} = 1.85 \,\, \Omega/\square$, Fermi energy $E_F = 12.8$ meV and $\ell_{\mathrm{MR}} = 92\,\mu\mathrm{m}$ which is amply long to place the measurements in the ballistic transport regime. Mesoscopic geometries were patterned by wet etching following electron-beam lithography. A magnetic field $B$ is applied perpendicular to the 2DES, allowing the formation of ballistic semiclassical electron cyclotron orbits of diameter 
$d_c= \frac{2 \hbar k_F}{e B}$, where $\hbar$ is the Planck constant, $e$ is the electron charge and $k_F= \sqrt{2\pi N_s} = 1.52 \times 10^8 \, \mathrm{m^{-1}}$ denotes the Fermi wavevector.

The device geometry of Fig.~\ref{fig1}(a)~\cite{Nihey_1990, F.Nihey_1990} consists of a total of 11 parallel barriers separated by 12 narrow conducting apertures. The center-to-center spacing between adjacent apertures is $L_c = 3\,\mu$m, with conducting aperture widths $W \, \approx \, 0.6\,\mu\mathrm{m}$ (taking into account a $\sim 0.1\,\mu$m depletion layer). As illustrated in the schematic in Fig.~\ref{fig1}(b), the multiple apertures in the structure can act simultaneously as electron injectors (solid arrows) and collectors (dashed arrows) of the ballistic semiclassical electron cyclotron orbits when $B$ is applied. The figure depicts only three parallel barriers for clarity, and shows schematic cyclotron orbits of different $d_c$. Here, $m$ denotes the number of barrier spacings between injector and collector (maximum 11), and $n$ denotes the number of $d_c$ fitting between injector and collector. Assuming the semiclassical electron trajectories follow cyclotron orbits, from Fig.~\ref{fig1}(b) the total length of the semiclassical trajectory is $(\pi/2) \, m L_{c}$, solely determined by $m$.

In the 4-point setup in Fig.~\ref{fig1}(a), using ac lock-in techniques, we measure the ac voltage drop $V_\mathrm{AC}$ over the perforated barrier under small applied ac current $I_\mathrm{AC}$ as a function of $B$. Because below we report on the effect of electron heating, which depends on the current through each identical single aperture, $I_\mathrm{AC}$ here denotes the current through a single aperture and is calculated as $\frac{1}{12}$ the total applied current. The measurements use $I_\mathrm{AC}$ = 8.33 nA per aperture corresponding to 100 nA total. The 4-point resistance of each individual aperture is then $R = V_\mathrm{AC}/I_\mathrm{AC}$, and is plotted versus $B$ to give the magnetoresistance $R(B)$ as shown in Fig.~\ref{fig1}(c).  The resistance of the entire perforated barrier is of course $\frac{1}{12}$ the resistance indicated in Fig.~\ref{fig1}(c). Symmetric maxima appear in $R(B)$, corresponding to ballistic TMF maxima~\cite{Adbhut2021}. Figure~\ref{fig1}(c) depicts the evolution of different TMF maxima dependent on $T$. Maxima in the $R(B)$ occur at values of $|B|$ such that $d_c = \frac{2 \hbar k_F}{e B} = \frac{m}{n} L_c$, which provides information about which trajectory, in terms of $m$ and $n$, corresponds to each focusing peak.

\section{Results and Discussion}

When electrons are injected from one injector aperture and the orbits impinge on the adjacent collector aperture such that $d_c = L_c = 3\,\mu\mathrm{m}$, the trajectory corresponds to the case $m=1$, $n=1$, depicted as the red cyclotron orbits in Fig.~\ref{fig1}(b) and corresponding to resistance peak \textbf{A} in Fig.~\ref{fig1}(c). When electrons reflect once from a barrier before reaching the adjacent collector such that $d_c = L_c/2 = 1.5\,\mu\mathrm{m}$, the trajectory corresponds to $m=1$, $n=2$ , depicted as the blue cyclotron orbits in Fig.~\ref{fig1}(b) and corresponding to resistance peak \textbf{B} in Fig.~\ref{fig1}(c). When electrons reflect twice from a single barrier before reaching the adjacent collector, the trajectory corresponds to $m=1$, $n=3$  (resistance peak \textbf{C}, omitted from the cyclotron orbit picture to reduce visual complexity but forms part of the discussion later). Other selected cyclotron orbits are also illustrated: for $m=2$, $n=1$ (peak \textbf{D}, purple) with $d_c = 2L_c = 6\,\mu\mathrm{m}$; and for $m=3$, $n=1$ (pink) with $d_c = 3L_c = 9\,\mu\mathrm{m}$, which is not labeled, as it will not be discussed further. In this work, we will mainly refer to the resistance peaks corresponding to trajectories \textbf{A}, \textbf{B}, \textbf{C} ($m=1$), and \textbf{D} ($m=2$) with the primary analysis based on the main resistance peak \textbf{A}  ($m=1$, $n=1$).

The use of the ballistic maxima in $R(B)$ is based on the sensible assumption that the amplitude of the maxima correlates with the number of electrons reaching the collector. In the following, the amplitude of the maxima was obtained by subtracting $R$ at the adjacent (higher $|B|$) minimum from $R$ at the corresponding maximum. We analyze the dependence on $T$ and on $I_{\mathrm{DC}}$ of the amplitudes of the maxima in $R(B)$. We start with the dependence on $T$, from which we determine $\ell_{\mathrm{MC}}(T)$. Apart from the behavior of the maxima, an analysis of $R$ at $B=0$ reveals an additional feature: with increasing temperature, $R(B=0)$ initially decreases and then increases, a phenomenon that will be discussed later. 

The amplitudes $A_\mathrm{m,n}$ of the maxima in $R(B)$ are lowered by scattering events, which are quantified by $\ell_{\mathrm{MR}}(T)$ and $\ell_{\mathrm{MC}}(T)$. Due to the dependence on $T$ of scattering events, the $A_\mathrm{m,n}$ exhibit a decay vs $T$, a behavior observed in earlier studies ~\cite{Nihey_1990, F.Nihey_1990, Spector_1990, Hornsey_1993, Adbhut2021, Egorov}. Both finite $\ell_{\mathrm{MR}}(T)$~\cite{Nihey_1990, F.Nihey_1990, Spector_1990, Hornsey_1993} and $\ell_{\mathrm{MC}}(T)$ can contribute to limiting $A_\mathrm{m,n}$. Yet, in the high-mobility GaAs 2DES featuring long $\ell_{\mathrm{MR}}(T)$, where momentum relaxation to the lattice by impurity scattering is weak, and at sufficiently low $T$ where electron-phonon scattering is weak ($T \lesssim 15$ K), the dominant scattering mechanism affecting $A_\mathrm{m,n}$ is e-e scattering ~\cite{Gurzhi, Scaffidi, Gupta_21, Egorov}, which lowers the number of individual electrons reaching the collector. The dependence on $T$ of $A_\mathrm{m,n}$ then enables us to extract $\ell_{\mathrm{MC}}(T)$ using the expression: 
\begin{equation}
A_\mathrm{m,n}(T) = A_0 S^{n-1} \\exp\left(-\frac{(\pi/2) \, m L_{c}}{\ell_{\mathrm{MC}}(T)}\right)
\label{eqamn}
\end{equation}
where $A_0$ is a proportionality constant for a given $N_s$, and the numerator in the exponential expresses the total length of the semiclassical trajectory. $S$ represents a coefficient capturing the decrease in amplitude for increasing number $n-1$ of reflections from the barriers. $S$ depends on the specularity for electron scattering from the barriers~\cite{Adbhut2021,keser2021,Nihey_1990,F.Nihey_1990}, and on the ability of electrons to focus on a collector aperture after scattering from the barrier, which can be influenced by the flatness of the barrier obtained by the fabrication process. For short $d_c$, $A_\mathrm{m,n}$ also depends on $d_c$ since the semiclassical trajectories are less well defined when $d_c$ approaches the value of $W$. Since, $d_c = \frac{m}{n}L_c$ we expect the effect of short $d_c$ to manifest as a decrease in $S$ for increasing $n$, feigning the effect of specularity. We estimate $S$ by analyzing the amplitude ratios of maxima with $m=1$ and increasing $n$, namely \textbf{A} ($n=1$), \textbf{B}  ($n=2$), and \textbf{C}  ($n=3$). This analysis yielded $S \approx 0.54 \pm 0.06$. The low value of $S<1$ indicates that specularity for scattering off the barriers is only one of several factors determining $S$. The wet etching method followed here indeed typically produces barriers of high specularity~\cite{Adbhut2021,F.Nihey_1990}. Due to the impact on $A_\mathrm{m,n}$ and $S$ of the mechanisms described above, this work relies mostly on peak \textbf{A} ($m=1,n=1$) for the determination of parameters, with peaks \textbf{B} and \textbf{C}  used for comparisons and supporting analysis. 

In accordance with phase-space arguments for the $T$ scaling of e-e interactions in Fermi liquid theory, \( \ell_\mathrm{MC} \propto \frac{1}{T^2} \)~\cite{Adbhut2021,narozhny2022hydrodynamic, Sarypov}. We thus express: 
\begin{equation}
    \ell_{\mathrm{MC}}(T) = \frac{\ell_0}{T^2}
\label{eqlmc}
\end{equation}
where $\ell_0$ is a constant. The literature contains expressions for the e-e scattering length featuring logarithmic corrections ~\cite{Giuliani,zheng1996, fu2018effect,Wang1, kim2020control}, which are omitted in Eq. \ref{eqlmc}. The logarithmic corrections originate in small-angle scattering events due to distant electrons and are relevant for the quantum lifetime as limited by e-e interactions ~\cite{nagaev2020, Giuliani, zheng1996, hofmann2023anomalously}. However, transport lifetimes are less affected by small-angle scattering and hence the logarithmic corrections can be omitted to a good approximation as in Eq. \ref{eqlmc} ~\cite{hofmann2023anomalously, Polini}. The use of $\ell_\mathrm{MC}(T)$ in Eq. \ref{eqamn} and in the analysis assumes that e-e scattering is dominant over electron-phonon scattering and hence assumes sufficiently low $T$. The analysis below shows that deviations appear in the range of $T$ where electron-phonon scattering cannot be neglected. The deviations are hence attributed to electron-phonon scattering. 


\begin{figure*}[htbp]
    \centering
    \includegraphics[width=\textwidth]{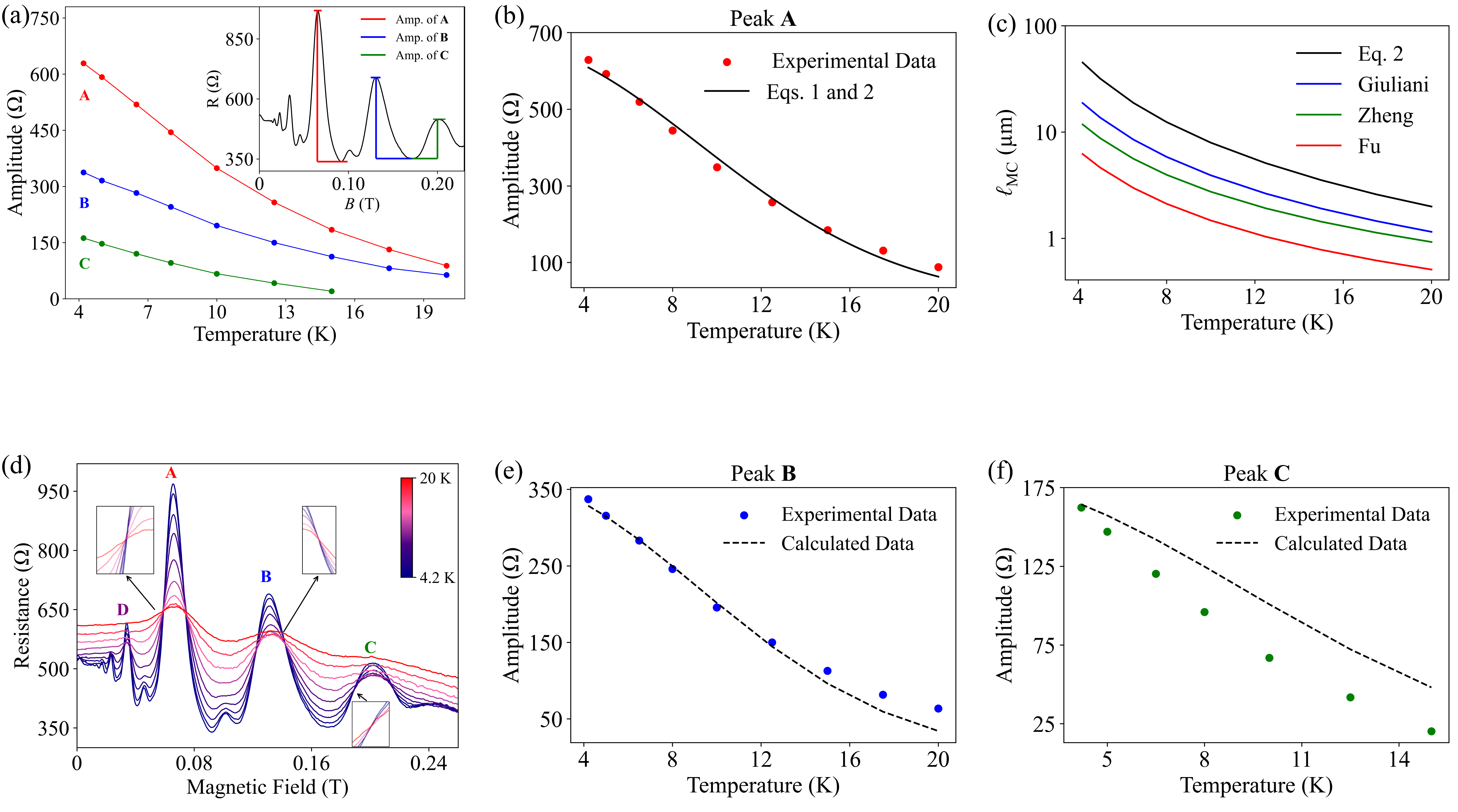}
    \caption{(a) Experimentally obtained amplitudes vs $T$ ($4.2 K \leq T \leq 20 K$) for peaks \textbf{A}, \textbf{B} and \textbf{C}. The inset shows the method for amplitude (Amp)
    quantification. (b) Fit of experimental amplitude for peak \textbf{A}  ($m$ = 1, $n$ = 1) by Eqs. \ref{eqamn} and \ref{eqlmc}. The red circles denote experimental data and the black line denotes the fit calculated using Eqs. \ref{eqamn} and \ref{eqlmc}. (c) $\ell_\mathrm{MC}$ vs $T$ plotted on a semi-logarithmic scale. Values calculated from Eq. \ref{eqlmc} are shown by the black line, where $\ell_{0}$ is extracted from the fit to experimental data in panel (b). The blue, green, and red lines represent theoretical predictions from Giuliani \cite{Giuliani}, Zheng \cite{zheng1996} and Fu \cite{fu2018effect}, respectively. (d) $R(B)$ vs $B$ and vs $d_c$ for peaks \textbf{A}, \textbf{B} and \textbf{C} corresponding to different $d_c$ and different $n$, showing invariant points. Each inset highlights the crossing of curves around the invariant point for that peak at different temperatures: peak \textbf{A} (left) up to $17.5$ K, peak \textbf{B}  (middle) up to $12.5 $ K, and peak \textbf{C}  (right) up to $8$ K. Arrows indicate the regions of the main panel that are magnified in the corresponding insets. (e,f) Comparison between experimentally measured and calculated amplitudes for higher-order peaks. Panels (e,f) show the amplitudes of peak \textbf{B}  ($m=1,n=2$) and peak \textbf{C} ($m=1,n=3$), respectively. Circles represent experimental data, while the dashed lines represent amplitudes calculated using Eqs. \ref{eqamn} and \ref{eqlmc} using parameters extracted from the fit to peak \textbf{A} and using $S=0.54$.}
    \label{fig2}
\end{figure*}

 In the following, we determine $\ell_\mathrm{MC}(T)$ primarily using the $T$ dependence of the main focusing peak \textbf{A} ($m=1,n= 1$), using Eqs. \ref{eqamn} and \ref{eqlmc}. To test the applicability of the same exponential model to other peaks, we then compare the parameters extracted from peak \textbf{A} with the peaks \textbf{B}($m=1,n=2$) and \textbf{C} ($m=1,n=3$).  No independent fitting was performed on peak \textbf{B} and \textbf{C}, and this comparison allows us to assess whether the parameters from peak \textbf{A} are consistent with these other peaks. In Fig.~\ref{fig2}(a), we plot the amplitude of $R$ corresponding to peaks \textbf{A}, \textbf{B}, and \textbf{C} of Fig.~\ref{fig1}(c) as a function of temperature $T$. The inset indicates how the amplitudes were determined for $T=4.2\,$K, as explained above. In the inset, the color of each marked peak matches the color of the data points in the main panel. The procedure in the inset for determining the amplitudes of the maxima was repeated for all measured $T$ of Fig. \ref{fig2}(a). Figure \ref{fig2}(b) shows the result of the fit to the amplitude corresponding to peak \textbf{A} vs $T$ using Eqs. \ref{eqamn} and \ref{eqlmc}. The fit yields $A_0 = 675.5~\Omega$ and $\ell_0 = 7.95\times 10^{-4}~\mathrm{m.K^2}$.
 The fitted curve (black line) closely follows the experimental data, indicating good agreement between model and measurement, supporting the role of e-e scattering in determining the amplitudes, and supporting the validity of Eqs. \ref{eqamn} and \ref{eqlmc} in the context of $R(B)$ in this multiparallel aperture geometry. Using the extracted $\ell_0$, we calculate values for $\ell_{\mathrm{MC}}$, as shown in Fig. \ref{fig2}(c). On the semilogarithmic scale, the calculated values and three theoretical predictions (\cite{zheng1996}, \cite{Giuliani}, \cite{fu2018effect}) exhibit a similar $T$ dependence over the measured range. This indicates that the functional form of the $T$ dependence is qualitatively well captured by the theories. However, the data calculated on the basis of the experiment lie at systematically higher $\ell_{\mathrm{MC}}$ than the theoretical estimates, differing mainly by a multiplicative scaling factor. Among the models considered,  across the entire $T$ range, the Giuliani theory \cite{Giuliani} yields the largest values of $\ell_{\mathrm{MC}}$, while the Fu model \cite{fu2018effect} predicts the shortest $\ell_{\mathrm{MC}}$. Discrepancies by a multiplicative factor have been noted previously between theoretical predictions for $\ell_{\mathrm{MC}}$ and values extracted from experiments in the ballistic or hydrodynamic transport regimes in 2DESs ~\cite{zheng1996, keser2021, Wang1, Egorov, Adbhut2021, Sarypov}. A discussion into the origins of the multiplicative factor has to await future developments.

Figure \ref{fig2}(d) shows $R(B)$ plotted vs both $B$ and $d_c$, for $B>0$ (same data as Fig. \ref{fig1} (c)). The figure and its insets indicate several invariant points through which all curves pass irrespective of $T$. A similar behavior was reported by Egorov \textit{et al.}~\cite{Egorov}, who attributed these invariant points to families of curves controlled by a single shared parameter. Based on their work and our observation of the role of e-e scattering in TMF, this parameter likely corresponds to e-e scattering, which governs the measured $R(B)$. The consistent crossing at invariant points fades at sufficiently high $T$, which we attribute to increased electron–phonon scattering challenging the dominance of e-e scattering. In our measurements, the invariant points persist up to different $T$ for different maxima, reflecting the onset of deviations from dominant e-e scattering: for peak \textbf{A} ($m=1, \, n=1$), all curves pass through the same invariant points except for $T = 20$ K (highest measured $T$). For peaks \textbf{B}  ($m=1, \, n=2$) and \textbf{C} ($m=1, \, n=3$), the invariant points persist up to $T \approx 12$ K, and $T \approx 8$ K, respectively, after which deviations appear. This trend indicates that different maxima and hence different semiclassical trajectories experience breakdown of the assumption of dominant e-e scattering at different $T$. We return to similar observations in the context of Figs.~\ref{fig2}(e, f). 

To assess the applicability of Eqs. \ref{eqamn} and \ref{eqlmc} to higher-order peaks, we extend the analysis to peak \textbf{B} ($n=2$, Fig.~\ref{fig2}(e)) and \textbf{C} ($n=3$, Fig.~\ref{fig2}(f)).  We extract the amplitudes of the peak \textbf{B} as shown in the inset of Fig.~\ref{fig2}(a). Yet, for peak \textbf{C}, the amplitude was referenced to the same minima used for peak \textbf{B}, namely the minima at lower $B$ relative to peak \textbf{C}. We did not use the adjacent minima at higher $B$ as used for peaks \textbf{A} and \textbf{B}, because those minima could not be resolved for $T > 5$~K. We use the previously extracted values of $A_0$ and $\ell_0$ (from fit to peak \textbf{A}), and additionally use $S=0.54$. Using these parameters, the expected amplitudes were calculated from  Eq. \ref{eqamn} and compared with the experimental data, as shown in Figs. \ref{fig2}(e, f).
\begin{figure*}[htbp]
    \centering
    \includegraphics[width=\textwidth]{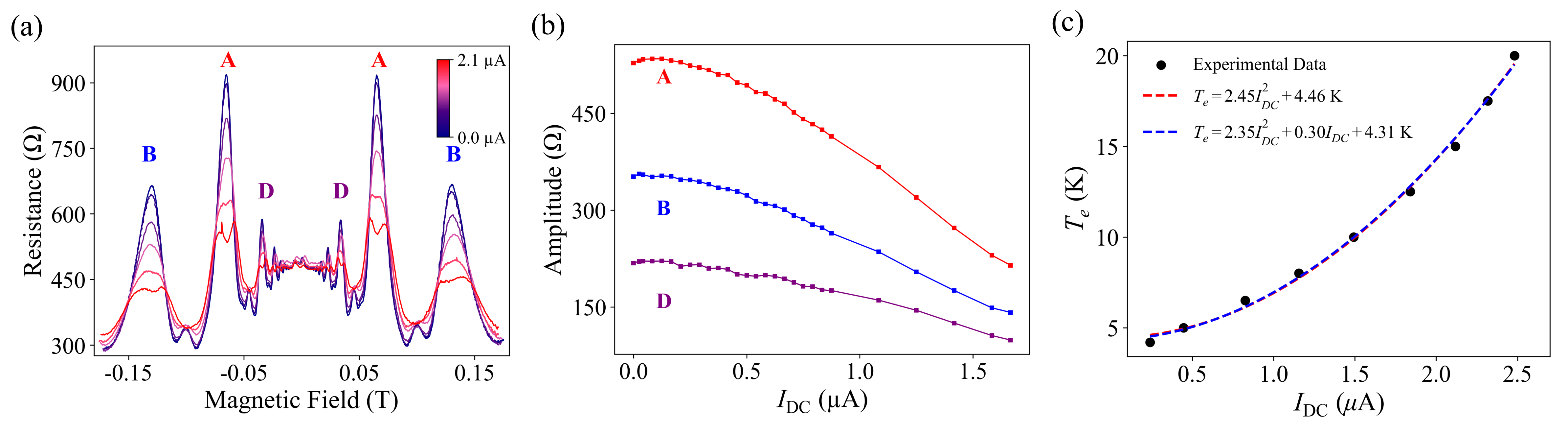}
    \caption{(a) Magnetoresistance measurements at $T$ = 4.2 K parametrized in $I_{\mathrm{DC}}$, showing maxima corresponding to different $m$ and $n$. Peak \textbf{A} corresponds to $m=1, \, n=1$; peak \textbf{B} to $m=1, \, n=2$; peak \textbf{D} to $m=2, \, n=1$. (b) Experimentally obtained amplitudes vs $I_{\mathrm{DC}}$ for peaks \textbf{A}, \textbf{B} and \textbf{D} at 4.2\,K. (c) $T_e$ vs $I_{\mathrm{DC}}$ derived from amplitude data for peak \textbf{A}. Black circles represent the data. The red and blue dotted lines represent fits without and with a linear term, respectively. The current and resistance values correspond to a single aperture.}
    \label{fig3}
\end{figure*}
For peak B (Fig.~\ref{fig2}(e)), the calculated amplitudes show good agreement with the experimental results up to approximately 12\,K, beyond which deviations become more pronounced. This increasing discrepancy at higher $T$ is likely again due to the breakdown of the assumption that the ballistic transport is limited solely by e-e scattering while at higher $T$ electron-phonon scattering becomes significant. For peak \textbf{C} (Fig.~\ref{fig2}(f)), the model of Eq. \ref{eqamn} underestimates the amplitude across most of the $T$ range, except at 4.2~K. Similar behavior has previously been reported, where the onset of electron-phonon scattering at higher $T$ led to the suppression of TMF signals~\cite{Krishna}. 

It is important to note that, unlike the semiclassical trajectory associated with peak \textbf{A} which follows a direct ballistic path between the adjacent injector and collector without boundary interaction, the trajectories for peaks \textbf{B} and \textbf{C} involve one and two reflections from the device boundary, respectively. The data shows that such boundary scattering introduces additional complexity that appears to affect the dependence on $T$ of the amplitudes, particularly at elevated $T$. In Fig.~\ref{fig2}(d) we observed that for peak \textbf{B} all curves cross at the same invariant points for $T \le 12$~K, while for $T > 12$~K deviations appear. The observation of the growing mismatch between calculated and experimental amplitudes and the breakdown of invariant points at similar $T$ may not be coincidental but are interpreted as two mutually consistent indicators of a scattering mechanism surpassing e-e scattering in limiting ballistic transport. As above, we surmise that this scattering mechanism is electron-phonon scattering. 

Peak \textbf{C} arises from a more complex ballistic trajectory involving two boundary scattering (reflection) events. Figure~\ref{fig2}(d) shows that curves for peak \textbf{C} pass through invariant points only up to 8 K, again suggesting that single-parameter behavior and adherence to Eqs. \ref{eqamn} and \ref{eqlmc}, which are determined by e-e scattering, are interrelated, and that effects of boundary scattering and electron-phonon scattering mutually interact. The physics of the mutual interaction is not presently clear and requires further study. 

We note that in ballistic transport similar behavior, where more intricate trajectories are more sensitive to $T$-dependent scattering, has been observed in other systems; for example, H. Chen \textit{et al.}~\cite{Chen2004} reported that higher-order ballistic $R(B)$ maxima in an InSb antidot lattice exhibited a distinct dependence on $T$ from maxima originating in simpler trajectories. Further, the Bloch-Gr\"{u}neisen temperature $T_\mathrm{BG} = 2 \hbar k_F s / k_B$ estimates the crossover temperature above which phonons with wave vectors $q = 2k_F$ become populated, hence when large-angle electron–phonon scattering becomes important \cite{kawamura1992phonon, SarmaPRB106-2022, stormer1990, raichev2017bloch, wang2}. Here, $k_B$ denotes the Boltzmann constant and $s$ the speed of sound in GaAs, with $s \, \approx$ 5000 m/s. We estimate $T_\mathrm{BG} \, \approx 12$ K. The coincidence of $T_\mathrm{BG}$ with the temperature scales where the deviations discussed above occur, suggests that large-angle electron–phonon scattering may be relevant both to the observed deviations from Eqs. \ref{eqamn} and \ref{eqlmc} of the amplitudes of peaks \textbf{B} and \textbf{C}, and to the deviations from single-parameter behavior. A detailed quantitative analysis of these effects is however beyond the scope of this work. 

We performed $R(B)$ measurements under a variable dc heating current $I_{\mathrm{DC}}$ at fixed $T=4.2$ K. In these measurements with $I_{\mathrm{DC}}\neq 0$, current heating causes $T_e>T$~\cite{Molenkamp,  Molenkamp_1994, DeJong, Gallagher, Starkov, Hornsey_1993,Rojek}, which allows a study of $T_e$ and its effects on ballistic transport. The local $T_e$ under current heating is well defined, since the e-e scattering time $\sim 10^{-12}$ s is far shorter than the electron-phonon scattering time $\sim 10^{-10}$ s \cite{Gallagher}. To measure $R(B)$, a small ac measurement current ($I_\mathrm{AC}$ = 8.33 nA per aperture corresponding to 100 nA total) was applied between the current contacts in Fig. \ref{fig1}(a). $I_{\mathrm{DC}}$ was superposed on $I_{\mathrm{AC}}$ to heat the 2DES through Joule heating \cite{Molenkamp, DeJong, Wang1, wang2, Molenkamp_Vol65}. In the following, $I_{\mathrm{DC}}$ denotes the dc current per aperture. As for the measurements at $I_{\mathrm{DC}} = 0$, $R(B)$ is obtained as $R(B) = V_\mathrm{AC}/I_\mathrm{AC}$, and hence closely approximates a differential resistance d$V$/d$I$. Figure~\ref{fig3}(a) displays the evolution of the $R(B)$ with increasing $I_{\mathrm{DC}}$, with the main peak \textbf{A}, and peaks \textbf{B} and \textbf{D} marked. At low $I_{\mathrm{DC}}$, several peaks corresponding to different $d_{c}$ can be clearly observed. As $I_{\mathrm{DC}}$ increases, the weaker peaks gradually vanish, consistent with the effects of increasing $T_e$. 

On increasing $I_{\mathrm{DC}}$, the more prominent peaks (\textbf{A}, \textbf{B}, \textbf{D}) begin to split. The splitting first appears for $I_{\mathrm{DC}} \approx 1.7\,\mu\text{A}$ and becomes more pronounced at higher $I_{\mathrm{DC}}$, as evidenced by the development of a minimum between the two closely spaced maxima. Following the explanation proposed by T. M. Chen \textit{et al.}~\cite{chen_2013}, the $I_{\mathrm{DC}}$-induced splitting can be understood as a consequence of the different electrochemical potentials on both sides of the perforated barrier under finite dc bias. The dc bias voltage $V_{\mathrm{DC}}=R(B) \, I_{\mathrm{DC}}$ shifts the equilibrium Fermi energy into two electrochemical potentials, one higher, at the source side in equilibrium with the electrochemical potential at the current source, $\mu_{\text{S}}$, and one lower, at the drain side in equilibrium with the electrochemical potential at the current drain, $\mu_{\text{D}}$. Since, $d_c$ depends on $k_F$, which in turn depends on the local electrochemical potential, $\mu_{\text{S}}$ and $\mu_{\text{D}}$ will correspond to different $d_c$ and hence produce TMF maxima at slightly different $B$. In a small-signal ac measurement our geometry yields TMF maxima corresponding to both $\mu_{\text{S}}$ and $\mu_{\text{D}}$ and hence produces two maxima closely spaced in $B$. As expected in this explanation, the splitting is more pronounced at higher $V_{\mathrm{DC}} \propto I_{\mathrm{DC}}$. 

The invariant points observed under variable $T$ in the $R(B)$ plots (Figs. \ref{fig1}(c) and \ref{fig2}(d)) are not present in the measurements under variable $I_{\mathrm{DC}}$ (Fig.~\ref{fig3}(a)). The amplitude of the main peak \textbf{A}, plotted in Fig.~\ref{fig3}(b) vs $I_{\mathrm{DC}}$, shows a non-monotonic dependence: it increases with $I_{\mathrm{DC}}$ at low $I_{\mathrm{DC}}$ and decreases at higher $I_{\mathrm{DC}}$. The decrease in amplitude for $I_{\mathrm{DC}} > 2\,\mu\mathrm{A}$ is expected and is attributed to enhanced e-e scattering caused by electron heating. However, the reason for the initial increase in amplitude for $I_{\mathrm{DC}} < 2\,\mu\mathrm{A}$ is not yet understood. The non-monotonic trend qualitatively resembles a transition from the ballistic Knudsen electron transport regime to the Gurzhi hydrodynamic regime~\cite{Molenkamp_1994, De_Jong}. A detailed quantitative analysis, while needed to unambiguously determine the transport regimes, lies outside the scope of this work. 

 In the experiments, we correlate $T_e$ and $I_{\mathrm{DC}}$ by comparing the amplitude $A_\mathrm{1,1}$ of peak \textbf{A} vs $T$ (Fig.~\ref{fig2}(a), red line and circles; Fig.~\ref{fig2}(b), red circles) to the same $A_\mathrm{1,1}$ vs $I_{\mathrm{DC}}$ (Fig.~\ref{fig3}(b), red line and squares). By finding the $T$ and $I_{\mathrm{DC}}$ where $A_{1,1}(T) = A_\mathrm{1,1}(I_{\mathrm{DC}})$, and identifying this $T$ with $T_e$ at the given $I_{\mathrm{DC}}$, we obtain the $T_e-I_{\mathrm{DC}}$ relation graphed in Fig.~\ref{fig3}(c) (black circles, experimental data). The method does not assume a model, but is fully experimental, being based on a direct comparison of differential resistance amplitudes obtained under variable $T$ and variable $I_{\mathrm{DC}}$. Figure~\ref{fig3}(c) (black circles) captures the effect of current heating on the 2DES. 

While the $T_e -I_{\mathrm{DC}}$ relation is obtained experimentally, applying $I_{\mathrm{DC}}$ both increases $T_e$ through Joule heating and injects electrons with an excess energy, $\Delta = eV_{\mathrm{DC}}=eR(B\approx0) \, I_{\mathrm{DC}}$. For the maximum current used in the measurements,
$I_{\mathrm{DC}}=2.1~\mu\mathrm{A}$ through each aperture, we estimate the corresponding $\Delta \, \approx 1.1~\mathrm{meV}$. Both the energy scales $k_BT_e$ and $\Delta$ can influence e-e scattering and consequently modify $\ell_{\mathrm{MC}}$. The dependence of $\ell_{\mathrm{MC}}$ on these energy scales can be obtained from the expressions given in Refs.~\cite{Giuliani, zheng1996}. After omitting the logarithmic correction terms, the expression involving $\Delta$ can be written as~\cite{linke1997, muller1995, Wang1, yacoby1991}
\begin{equation}
\frac{1}{\ell_{\mathrm{MC}}}=
\frac{E_{\mathrm{F}}}{\alpha \hbar v_{\mathrm{F}}}
\left(\frac{\Delta}{E_{\mathrm{F}}}\right)^2 ,
\label{eq:Delta}
\end{equation}
and the corresponding expression involving Joule heating and $k_BT_e$ can be written as~\cite{De_Jong, Molenkamp, Molenkamp_1994, Molenkamp_Vol65,Adbhut2021}
\begin{equation}
\frac{1}{\ell_{\mathrm{MC}}}
=
\frac{E_{\mathrm{F}}}{\beta \hbar v_{\mathrm{F}}}
\left(\frac{k_{\mathrm{B}}T_e}{E_{\mathrm{F}}}\right)^2 ,
\label{eq:Joule_heating}
\end{equation}
where $v_F$ is the Fermi velocity, and $\alpha$ and $\beta$ are parameters. Equations~(\ref{eq:Delta}) and~(\ref{eq:Joule_heating}) show that $\Delta$ and $k_BT_e$ enter $\ell_{\mathrm{MC}}$ through the same quadratic dependence on the corresponding energy scale. Since the applied $I_{\mathrm{DC}}$ simultaneously produces Joule heating and $\Delta$, their individual contributions to $\ell_{\mathrm{MC}}$ cannot be separated directly in the experiment. The measured dependence of $\ell_{\mathrm{MC}}$ on $I_{\mathrm{DC}}$ therefore reflects the combined effect of these two energy scales.  

To describe the contribution of Joule heating to the increase in $T_e$, we follow the discussion in Refs. \cite{Molenkamp, Molenkamp_1994, Molenkamp_1991, Molenkamp_Vol65, Rojek}, where the local $T_e$ is determined by a balance between two processes: the current flowing through the device produces Joule heating of the 2DES, while the electrons lose energy to the lattice which acts as a thermal reservoir at fixed lattice temperature $T_L$. The competition between Joule heating and electron-lattice relaxation leads to a steady state heat-balance condition, expressed as: 
\begin{equation}
C_v \,(T_{e}-T_{L}) = \left(\frac{I_{\mathrm{DC}}}{W}\right)^2 \, R_{\square} \, \tau_\varepsilon
\label{eq:heat_balance}
\end{equation}
where $R_{\square}$ denotes the 2D resistivity, $\tau_\varepsilon$ the electron energy relaxation time characterizing the transfer of energy from the 2DES to the lattice by phonon scattering, and $C_v$, the 2D heat capacity of the 2DES ~\cite{Rojek, Molenkamp_1991}. In the present work, the extracted $\tau_\varepsilon$ from Eq.~\ref{eq:heat_balance} remains on the order of $\sim 10^{-11}~\mathrm{s}$ over the measured $T_e$ range, showing little variation and remaining consistent with previous reports~\cite{Molenkamp}. This heat-balance equation yields a quadratic dependence of $T_e$ increase on the applied $I_{\mathrm{DC}}$, $T_e-T_L \propto I_{\mathrm{DC}}^2$, as expected for a Joule heating model.  

As discussed, $I_\mathrm{{DC}}$ also produces an excess injection energy $\Delta$. Equations~\ref{eq:Delta} and \ref{eq:Joule_heating} show that $k_BT_e$ and $\Delta$ enter $\ell_{\mathrm{MC}}$ with the same functional dependence. Consequently in the absence of Joule heating, $\Delta$ corresponds to an effective thermal energy scale, $k_BT_e\sim \Delta$. Since $\Delta$ is linear in $I_{\mathrm{DC}}$, the corresponding effective $k_BT_e$ would also increase linearly with $I_{\mathrm{DC}}$, in contrast to the quadratic dependence expected from Joule heating.  

The experimentally extracted $T_e-I_{\mathrm{DC}}$ relation shown in Fig.~\ref{fig3}(c) can therefore be used to assess the relative contributions of Joule heating and excess injection energy. In Fig.~\ref{fig3}(c), the extracted $T_e-I_{\mathrm{DC}}$ relation was fitted using both a quadratic dependence and a combined quadratic plus linear dependence. A quadratic fit yields
$T_e = 2.45\,\mathrm{\frac{K}{\mu A^2}}\,I_{\mathrm{DC}}^2 + 4.46\,\mathrm{K}\,$
and the fit including an additional linear term yields
$T_e = 2.35\,\mathrm{\frac{K}{\mu A^2}}\,I_{\mathrm{DC}}^2 + 0.30\,\mathrm{\frac{K}{\mu A}}\,I_{\mathrm{DC}} + 4.31\,\mathrm{K}$ with $I_{\mathrm{DC}}$ expressed in $\mu$A. The $T_e-I_{\mathrm{DC}}$ relation closely follows the $I_{\mathrm{DC}}^2$ dependence predicted by Eq.~\ref{eq:heat_balance}. Although the fit including a linear term yields a non-zero linear coefficient, the contribution of the linear term to the overall increase in $T_e$ is small compared with the quadratic term. We therefore conclude that the increase in $T_e$ is dominated by the quadratic current dependence expected for Joule heating, consistent with earlier studies of current-induced electron heating~\cite{Molenkamp,Molenkamp_Vol65}. The quadratic dependence of $T_e$ on $I_{\mathrm{DC}}$ is also consistent with previous reports of current-induced cooling in 2DESs under $B$ field~\cite{cooling_Naomi}, where the dominant Joule heating contribution appears as \( \nabla^2 T \propto -R_{\square} J^2 / \kappa \), with $J$ the current density and $\kappa$ the thermal conductivity. 

However, linear and sublinear current dependences have also been reported in the literature. A sublinear dependence of $T_e$ on $I_{\mathrm{DC}}$ was reported in Ref.~\cite{Wang1}, where \(
T_e(I_{\mathrm{DC}})=T_0+1.0\,I_{\mathrm{DC}}^{0.65},
\)
with $T_0=0.04\,\mathrm{K}$, while Ref.~\cite{yacoby1991} reported an approximately linear dependence of $T_e$ on $I_{\mathrm{DC}}$. The different current dependences may be related to differences in device geometry. For modifying $T_e$ through $I_{\mathrm{DC}}$, a short channel or wire with an Ohmic voltage drop may not be equivalent to a structure containing a local potential barrier, where the voltage drop is spatially localized and the contribution of excess energy injection can be different. The possible influence of geometry on these effects is beyond the scope of the present work.  

Equation \ref{eq:heat_balance} hence captures the relation between $T_e$ and $I_{\mathrm{DC}}$ well. Even at zero $I_\mathrm{DC}$, the offset ($T_e \approx$~4.46\,K) shows that the electrons are slightly hotter than the lattice ($T_L$= 4.2\,K) due to imperfect thermal equilibration. Because the electron-phonon coupling is weak at lower $T$, the electrons do not fully equilibrate with the lattice and can maintain a temperature slightly above $T_L$. In short, by correlating the $R(B)$ peak amplitude with both $T$ and $I_{\mathrm{DC}}$, we were able to deduce a quantitative relation between $T_e$ and $I_{\mathrm{DC}}$. This shows that $I_{\mathrm{DC}}$ acts predominantly as a heating source, allowing deterministic control of $T_e$ in the mesoscopic environment and consequently allowing predictable modulation of e-e interactions. 

\begin{figure}[htbp]
    \centering
    \includegraphics[width=\columnwidth]{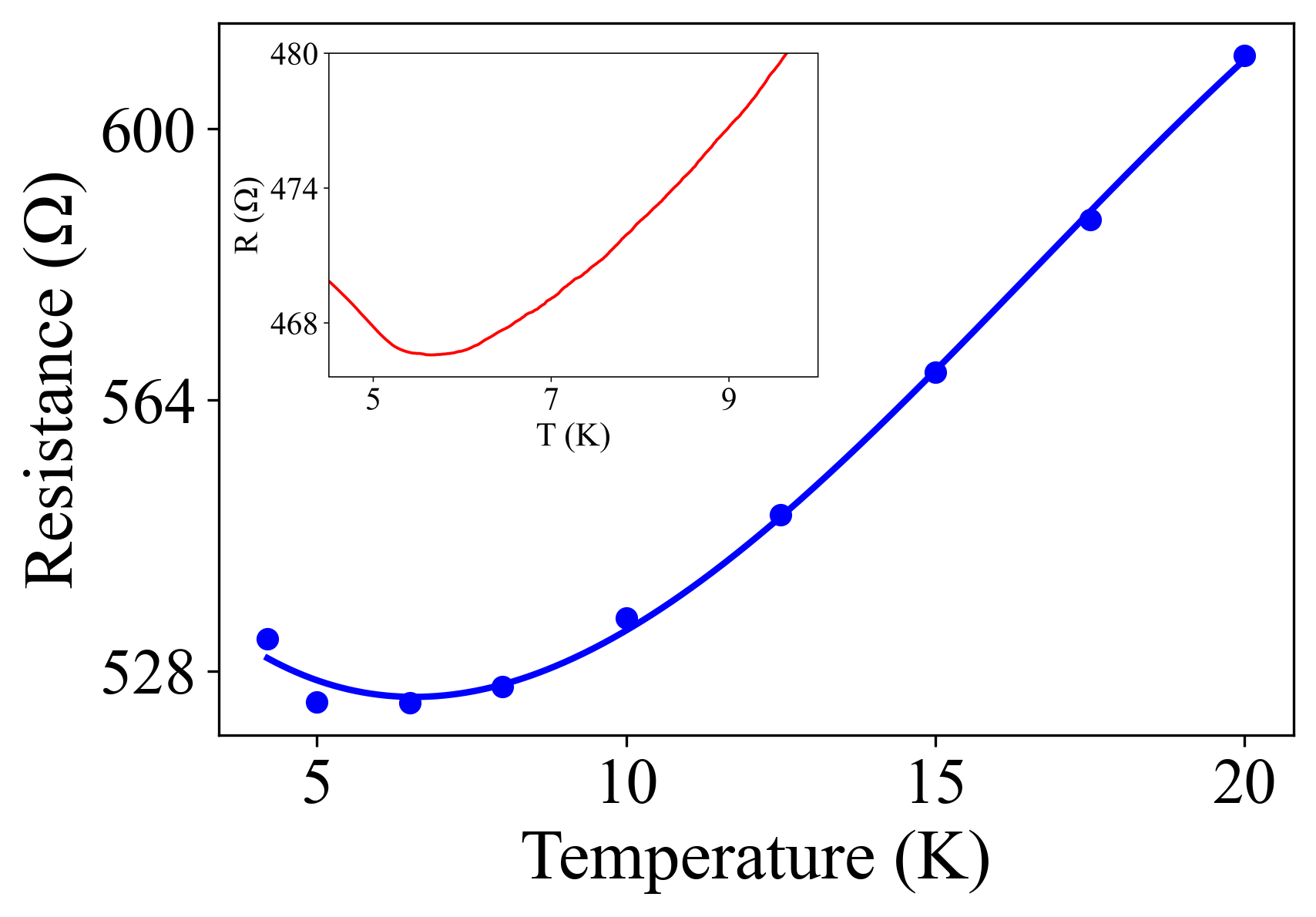}
    \caption{$R(B=0)$ vs $T$. The circles correspond to data obtained at $B=0$ from Fig.~\ref{fig1}(c) across $T$, while the line is a spline fit as guide to the eye. The minimum in $R$ vs $T$ is characteristic of the Gurzhi effect and the phenomenon of superballistic conduction. The inset shows corroborating $R(B=0)$ vs $T$ data recorded during warmup of the device. }

    \label{fig4}
\end{figure}

The perforated barrier geometry, since it consists of parallel narrow channels, is expected to show the Gurzhi electron hydrodynamic effect and the corresponding phenomenon of superballistic conduction ~\cite{Estrada-Alvarez, Krishna, Kravtsov}. In this hydrodynamic regime, enhanced e–e scattering suppresses the boundary resistance, giving rise to channel resistances below the ballistic limit. To investigate the Gurzhi effect and superballistic conduction~\cite{Estrada-Alvarez, Krishna, Kravtsov} in the perforated barrier geometry, in Fig.~\ref{fig4} we plot $R$ vs $T$ at $B$ = 0. The data for $R(B=0)$ are obtained from Fig.~\ref{fig1}(c) across a range of $T$ (circles in Fig.~\ref{fig4}). Figure~\ref{fig4} shows a decrease in $R$ with increasing $T$ up to approximately 6\,K, consistent with the Gurzhi effect. The minimum in $R$ vs $T$ indicates that superballistic conduction is indeed present, up to $T \approx 6$ K. The increase in $R$ beyond $\sim 6$ K marks the departure from the superballistic regime due to the dominance of momentum non-conserving electron–phonon scattering and indicates a gradual transition to the diffusive transport regime. Additional data of $R(B=0)$ vs $T$ obtained during a non-controlled warmup of the device are shown in the inset, confirming the same overall behavior. 

\section{Conclusion}
In conclusion, a multiparallel aperture geometry offers a compact platform for studying ballistic transport in two-dimensional electron systems. Analysis of amplitudes of maxima in ballistic magnetotransport  provides access to the electron–electron scattering length and electron temperature, both of which are key parameters in solid-state systems.  The amplitude of the maxima is observed to decay exponentially with temperature, enabling determination of the electron-electron scattering length. We further observe that different maxima deviate from their invariant points at different temperatures, showing that each semiclassical trajectory undergoes the breakdown of dominant electron-electron scattering independently. We also demonstrate that electron heating under dc bias follows a quadratic current dependence, consistent with Joule heating, allowing the electron temperature to be driven systematically away from the lattice temperature. This establishes electron temperature as a controllable parameter for exploring ballistic transport and non-equilibrium regimes in two-dimensional electron systems. We further observe signatures of hydrodynamic superballistic conduction. Collectively, these results highlight the rich tunability and unique transport phenomena enabled by the very high electron mobility in GaAs/AlGaAs paired with the mesoscopic multiparallel aperture geometry, from studies of electron–electron scattering and controlled electron heating to superballistic conduction.

\begin{acknowledgments}
We would like to thank Victoria Soghomonian and Manichandra Morampudi for suggestions and insightful discussions. The Princeton University portion of this research is funded in part by the Gordon and Betty Moore Foundation’s EPiQS Initiative, Grant GBMF9615.01 to L. N. Pfeiffer.
\end{acknowledgments}



\bibliography{PRB_reference}

\end{document}